\documentclass[reqno,11pt]{amsart}
\usepackage{epsfig}\usepackage{amsfonts,amsmath,amssymb,mathrsfs,verbatim,amsthm,amscd}
\newcommand{\beq}{\begin{equation}}
\newcommand{\ee}{\end{equation}}
\newcommand{\bea}{\begin{eqnarray}}
\newcommand{\eea}{\end{eqnarray}}

\usepackage{hyperref}

\newcommand{\p}{\partial_x}
\newcommand\id{1\kern-0.25em\text{l}}
\newcommand\zero{0\kern-0.4em\text{0}}

\newcommand{\R}{\mathbb R}

\newcommand{\T}{\mathbb T}

\newcommand{\be}{\begin{equation}}
\newcommand{\ba}{\begin{eqnarray}}
\newcommand{\ea}{\end{eqnarray}}

\begin{document}

\title[Variations of supersymmetric quantum mechanics]
{Variations of supersymmetric quantum mechanics \\
and  superconformal indices }

\author{Vyacheslav \,P. Spiridonov}%

\address{Laboratory of Theoretical Physics,
JINR, Dubna, Moscow region, 141980 Russia and
National Research University Higher School of Economics, Moscow, Russia
}

\maketitle

\vspace*{-2em}

\begin{abstract}
Old studies on supersymmetric quantum mechanics and its deformations,
that were initiated by the 1988 joint paper with V. Rubakov,
are retrospectively discussed. In the modern circumstances, corresponding results can be related
to computations of superconformal indices and associated special functions.
\end{abstract}

\begin{flushright}
\em To the memory of Valery Rubakov
\end{flushright}

\vspace*{1em}

Supersymmetry is hypothetically valid for quantum field theory of elementary particles,
but it has not found experimental confirmations yet. Nevertheless, a realization of supersymmetry
in quantum mechanical systems \cite{Wi81} has shown that this is not an abstract notion,
but a valid mathematical construction describing properties of real physical models.
This application triggered intense research of corresponding simple supersymmetric systems.
The joint work with V. Rubakov \cite{RS} appeared from a wish to extend the quantum field theory of bosons and
fermions to the parastatistical degrees of freedom. It was inspired by the question on the
possibility to break the Pauli principle by admitting the third allowed state for electron,
which was discussed at that time. The things appeared to be complicated and,
instead of the quantum field theory modifications, a parafermionic extension
of the supersymmetric quantum mechanics was proposed.

The standard supersymmetric quantum mechanics (SQM) is based on the following algebra
\begin{equation}\label{susy}
\{Q^+,Q^-\}=H, \qquad [H,Q^\pm]=(Q^\pm)^2=0,
\end{equation}
where $H$ is the Hamiltonian and $Q^\pm$ are conserved supercharges related by hermitian
conjugation, $(Q^-)^\dag =Q^+$. It is assumed that all operators are well defined in
the Hilbert space of physical states.
Equivalently, one can use the hermitian supercharges $Q_1=Q^++Q^-, Q_2=(Q^+-Q^-)/i$ and
the algebra takes the form
$
\{Q_j,Q_k\}=2H\delta_{jk},\; [H,Q_j]=0,\, j, k=1,2.
$

From this superalgebra it immediately follows that
the spectrum of the Hamiltonian must be semipositive, and all positive energy states
are doubly degenerate. Breaking of supersymmetry is determined by the existence of
the zero energy states vanishing under the action of supercharges $H|0\rangle=Q^{\pm}|0\rangle=0$.
The Witten index $I_W=Tr\,(-1)^Fe^{-\beta H}$, where $F$ is the fermion charge, equals to the difference
between the number of bosonic and fermionic vacua. When $I_W \neq 0$, supersymmetry
is definitely not broken. For simplicity we consider only systems with the discrete
spectrum, when there is no problem with the continuous spectrum going down to the
zero energy. Also it is convenient to assume that we deal with the Hilbert space $\textrm{L}^2(\R)$
in order to simplify the boundary conditions.

The simplest realization of this algebra uses the supercharges
\begin{equation}
Q^+=\left(\begin{array}{cc}
 0 & A^+ \\
 0   & 0
\end{array}\right),\quad
Q^-=\left(\begin{array}{cc}
 0 & 0 \\
 A^-   & 0
\end{array}\right),
\quad A^{\pm}=\mp \p +v(x),\quad \p:=\frac{d}{dx},
\label{SUSYchar}\end{equation}
which yield the Hamiltonian
\begin{equation}
H=\left(\begin{array}{cc}
h_1 & 0 \\   % L_1-\lambda_1
 0   &  h_2  % L_2-\lambda_2
\end{array}\right)=-\p^2+v^2(x)-v'(x)\sigma_3,
\quad \sigma_3=\left(\begin{array}{cc}
1 & 0\\
 0   & -1
\end{array}\right),
\label{HamSUSY}\end{equation}
where $v'(x)\equiv \p v(x)$.
Physically, one has a spin 1/2 particle on the line in an external magnetic field
in the vertical direction.

Consider this model in a more general context.
Take an infinite sequence of one-dimensional
Schr\"odinger operators and eigenvalue problems for them
\beq
L_j=-\partial_x^2 +u_j(x), \qquad L_j\psi^{(j)}(x)=\lambda \psi^{(j)}(x), \quad j\in \mathbb{Z}.
\label{DTL}\ee
Let us demand that the neighboring eigenfunctions are connected to each other by the action
of differential operators of the first order
\beq
\psi^{(j+1)}(x)=A_j^-\psi^{(j)}(x), \qquad A_j^-=\partial_x+v_j(x).
\label{DTE}\ee
The compatibility condition of equations in \eqref{DTL} and \eqref{DTE}
leads to the intertwining relation $L_{j+1}A_j^-=A_j^-L_j$.
Resolving it, one finds explicit connection between $u_j(x)$ and $v_j(x)$:
\beq
u_j(x)=v_j^2(x)-v_j'(x)+\lambda_j, \qquad u_{j+1}(x)=u_j(x)+2v_j'(x),
\label{ujfj}\ee
where $\lambda_j$ are the integration constants. Equivalently,
\beq
 v_{j+1}^2(x)-v_{j+1}'(x)+\lambda_{j+1}=v_j^2(x)+v_j'(x)+\lambda_j.
\label{chain}\ee

In the operator language one comes to the  factorization of the Schr\"odinger operators
\beq
L_j=A_j^+A_j^-+\lambda_j, \qquad L_{j+1}=A_j^-A_j^+ +\lambda_j=A_{j+1}^+A_{j+1}^-+\lambda_{j+1},
\label{fact}\ee
where
$
A_j^+= -\partial_x+v_j(x).
$
The hermitian conjugation conditions $A_j^+=(A_j^-)^\dag,\, L_j^\dag=L_j$ assume that these operators
are well defined on a sufficiently dense domain of L$^2(\R)$, in particular,
that all superpotentials  $v_j(x)$ do not have singularities spoiling
normalizability of $\psi_j(x)$ eigenfunctions.

The Lax pair (in the terminology of the theory of integrable systems)  \eqref{DTL} and \eqref{DTE}
was introduced by Infeld \cite{Inf}, but the transformation  \eqref{DTE} for an ordinary
differential equation of the second order was considered much earlier by Darboux.
Described formulas define basics of the factorization method for solving
eigenvalue problems in quantum mechanics initiated by Schr\"odinger himself,
see the survey \cite{IH}. Evidently, any pair of neighbouring operators $L_j$ can be used for
constructing the supersymmetric Hamiltonian \eqref{HamSUSY}, $h_1=L_{j}-\lambda_j, \; h_2=L_{j+1}-\lambda_{j}$.

The system  \eqref{SUSYchar}, \eqref{HamSUSY} contains a bosonic degree of freedom
described by the variables $x$ and $p=-i\p$, $[x,p]=i$, and a fermionic one
described by the matrices
$\displaystyle
f^+=\left(\begin{array}{cc}
 0 & 1\\
 0   & 0
\end{array}\right),\;
f^-=\left(\begin{array}{cc}
 0 & 0 \\
 1   & 0
\end{array}\right),
$
so that $(f^-)^2=(f^+)^2=0, \{f^-,f^+\}=1$.
Supercharges are built as products of bosonic and fermionic operators $Q^+=A^+f^+,\, Q^-=A^-f^-$.
Working with a bigger number of bosonic and fermionic degrees of freedom,
it is possible to build models with the extended supersymmetry
involving bigger number of conserved supercharges.

There are  statistics other than the bosonic and fermionic ones called parastatistics.
They describe different types of symmetrization or antisymmetrization of
wave functions for a system of identical particles. The latter are characterized by an integer $p$ describing
the number of rows (for parabosons) or columns (for parafermions) in the corresponding Young
diagrams. Parabosonic and parafermionic creation and annihilation operators satisfy some
general trilinear relations.
A variation of SQM proposed in \cite{RS} used the parafermion of order $p=2$ instead of the fermion.
It was called the parasupersymmetric quantum mechanics (PSQM).
The corresponding creation and annihilation operators satisfy relations
$$
a^3=0, \quad a^2a^+ + a^+a^2= 2a,\quad  aa^+a =2a
$$
and their hermitian conjugates following from relations $a^\dag=a^+,\, (a^+)^\dag =a$.
Operators $a, a^+$ can be realized by $3\times 3$ matrices which are not described here.
With their help, the following parafermionic (of order two) generalization of
supercharges was suggested in \cite{RS}
$$
Q^+=\left(\begin{array}{ccc}
 0 & A_1^+ & 0 \\
 0   & 0 & A_2^+ \\
 0 & 0& 0
\end{array}\right),\quad
Q^-=\left(\begin{array}{ccc}
 0 & 0 & 0 \\
 A_1^-  & 0 & 0 \\
 0 & A_2^- & 0
\end{array}\right), \quad (Q^\pm)^3=0.
$$
These operators generate the following parasuperalgebra
\begin{eqnarray}\nonumber &&
(Q^-)^2Q^+ + Q^-Q^+Q^- +Q^+(Q^-)^2=2Q^-H, \quad
\\  &&
(Q^+)^2Q^- + Q^+Q^-Q^+ +Q^-(Q^+)^2=2Q^+H,
\label{PSUSYalg1}\end{eqnarray}
where the Hamiltonian $H$ commutes with the parasupercharges, $[H,Q^\pm]=0$, and has
a $3\times 3$ diagonal matrix form
$$
H=-\p^2 +\textrm{diag}(v_1^2-v_1'-c, v_1^2+v_1' -c , v_2^2+v_2'+c)
$$
with arbitrary real constant $c$. The middle element of the Hamiltonian can be written
in the form $-\p^2+v_2^2-v_2'+c$
% due to the basic equation for functions $v_{1,2}(x)$:
because the functions $v_1(x)$ and $v_2(x)$ are connected by the differential equation
\begin{equation}
v_1'(x)+v_2'(x)+v_1^2(x)-v_2^2(x)=\lambda_2-\lambda_1 \equiv 2c.
\label{cons}\end{equation}
In terms of the Hamitonians $L_j$ defined earlier, we have
$$
H=\textup{diag} (L_1-a,L_2-a,L_3-a),\qquad a =\tfrac12 (\lambda_1+\lambda_2).
$$
Our notation differs from the one used in \cite{RS} by the changes $v_j(x)\to -W_j(x)$
and $2c\to -c$. We shall assume below that $c> 0$. The models with $c=0$ are too specific
for our goals and the case $c<0$ can be recovered by the changes $v_j(x)\to -v_j(x)$.

For the hermitian charges $Q_1=Q^++Q^-, Q_2=(Q^+-Q^-)/i$, one has the relations
$$
Q_i(\{Q_j,Q_k\}-2H\delta_{jk})+\mbox{cyclic perm. of }i,j,k  =0,
\qquad [H,Q_i]=0,
$$
The spectrum of $H$ is now triply degenerate with possible
exception of two smallest eigenvalues.

It is possible to realize the superalgebra \eqref{susy} using matrices \eqref{SUSYchar}
with $A^\pm$ given by a linear differential operator with $v(x)$ being a $2\times 2$ matrix superpotential.
In this case supersymmetric Hamiltonian $H$ is given by a $4\times 4$ matrix.
By appropriate choice of the elements of the matrix superpotential it is possible to diagonalize
the $2\times 2$ analogue of $h_1$ (or $h_2$) in \eqref{HamSUSY}. The Hamiltonian takes then
a block-diagonal form. As shown in \cite{AISV}, after deleting one $1\times1$ dimensional block
(i.e.,  one row and one column) in  $H$ there emerges a $3\times 3$ matrix Hamiltonian
satisfying the parasupersymmetric algebra. Its further simplification yields the above
diagonal PSQM Hamiltonian. Thus parasupersymmetric systems can be obtained by projecting (truncating) the
Hilbert spaces of supersymmetric systems to lower dimension subspaces. Sometimes this is a
natural procedure, if the truncated subspace subhamiltonian is not self-adjoint due to
some singularities of the potential.

Another type of variation of SQM was suggested
by Andrianov, Ioffe and the author in \cite{AIS}.
It was called the higher-derivative supersymmetric quantum mechanics (HSQM),
since it used  higher order differential operators as supercharges.  Consider the simplest case
when the supercharges are defined by matrix differential operators of the second order
\begin{equation}
Q^+=\left(\begin{array}{cc}
 0 & A_1^+A_2^+ \\
 0   & 0
\end{array}\right), \quad
Q^-=\left(\begin{array}{cc}
 0 & 0 \\
A_2^-A_1^-  & 0
\end{array}\right).
\label{HSUSYch}\end{equation}
They define the superalgebra
$$
\{Q^+, Q^-\}= (H-c)(H+c), \qquad [H, Q^\pm]=(Q^\pm)^2=0, \quad 2c=\lambda_2-\lambda_1,
$$
where the Hamiltonian has the form
\begin{equation}
H=\left(\begin{array}{cc}
-\p^2+v_1^2-v_1'-c & 0  \\
 0 & -\p^2+v_2^2+v_2'+c
\end{array}\right) =
\left(\begin{array}{cc}
L_1-a & 0  \\
 0 & L_3-a
\end{array}\right).
\label{HamHSQM}\end{equation}
Clearly, this Hamiltonian is obtained from the parasupersymmetric one by deleting the
middle subhamiltonian corresponding to a truncation of the space of states.
If one replaces in HSQM supercharges \eqref{HSUSYch} the product $A_1^+A_2^+$ by $A_1^+\cdots A_n^+$
and $A_2^-A_1^-$ by $A_n^-\cdots A_1^-$, then we have the general polynomial superalgebra
$$
\{Q^+, Q^-\}=\prod_{k=1}^n (H-\lambda_k), \qquad [H, Q^\pm]=(Q^\pm)^2=0, \quad
H=\left(\begin{array}{cc}
L_1 & 0  \\
 0 & L_{n+1}
\end{array}\right).
$$
In our $n=2$ case we shifted the Hamiltonian by a constant in order to match with the
parasupersymmetric case and with a different variation of SQM model considered below.
For a detailed description of various HSQM models and their properties, see \cite{AI}.

An interesting model of ``weak'' supersymmetric quantum mechanics (WSQM) was suggested by Smilga
in \cite{Sweak}. It represents further modification of PSQM and HSQM quantum mechanical models with
a non-standard superalgebra of symmetries.
Define the supercharges
$$
Q_1^-=\left(\begin{array}{cccc}
 0 & 0 & 0 & 0\\
A_1^- & 0 & 0 & 0 \\
A_1^- & 0 & 0 & 0 \\
0 & A_2^- & -A_2^- & 0 \\
\end{array}\right),
\quad
Q_2^-=\left(\begin{array}{cccc}
 0 & 0 & 0 & 0\\
-A_1^- & 0 & 0 & 0 \\
A_1^- & 0 & 0 & 0 \\
0 & A_2^- & A_2^- & 0 \\
\end{array}\right)
$$
and their hermitian conjugates $Q_\alpha^+=(Q_\alpha^-)^\dag$,
$$
Q_1^+=\left(\begin{array}{cccc}
 0 & A_1^+ & A_1^+ & 0\\
 0 & 0 & 0 & A_2^+ \\
 0 & 0 & 0 & -A_2^+ \\
0 & 0 & 0 & 0 \\
\end{array}\right),
\quad
Q_2^+=\left(\begin{array}{cccc}
 0 & -A_1^+ & A_1^+ & 0\\
 0 & 0 & 0 & A_2^+  \\
 0 & 0 & 0 & A_2^+ \\
0 & 0 & 0 & 0 \\
\end{array}\right).
$$

These supercharges satisfy the following algebraic relations
\begin{equation}
\{Q_\alpha^\pm,Q_\beta^\pm\}=0,\quad \{Q_\alpha^-,Q_\beta^+\}=2\big((H-Y)\delta_{\alpha\beta}+Z_{\alpha\beta}\big),
\label{WSQM}\end{equation}
where $H$ is the Hamiltonian given by the following $4\times 4$ diagonal matrix
\begin{equation}
H=-\p^2+\textup{diag} (v_1^2-v_1'-c,v_1^2+v_1'-c,v_2^2-v_2'+c,v_2^2+v_2'+c),
\label{ham_wsqm}\end{equation}
and
\begin{eqnarray*} &&  \makebox[0.7em]{}
Y=c\left(\begin{array}{cccc}
 -1 & 0 & 0 & 0\\
0 & 0 & 0 & 0 \\
0 & 0 & 0 & 0 \\
0 & 0 & 0 & 1 \\
\end{array}\right),
\quad
Z_{11}=c\left(\begin{array}{cccc}
0 & 0 & 0 & 0\\
0 & 0 & 1 & 0 \\
0 & 1 & 0 & 0 \\
0 & 0 & 0 & 0 \\
\end{array}\right)=-Z_{22},
 \\  &&
Z_{12}=c\left(\begin{array}{cccc}
0 & 0 & 0 & 0\\
0 & -1 & 1 & 0 \\
0 & -1 & 1 & 0 \\
0 & 0 & 0 & 0 \\
\end{array}\right),
\quad Z_{21}=Z_{12}^\dag=
c\left(\begin{array}{cccc}
0 & 0 & 0 & 0\\
0 & -1 & -1 & 0 \\
0 & 1 & 1 & 0 \\
0 & 0 & 0 & 0 \\
\end{array}\right).
\end{eqnarray*}
One has $[Q_\alpha^\pm, Y]=\pm c Q_\alpha^\pm$ and the Hamiltonian $H$, $Y$ and $Z_{\alpha\beta}$ commute with each other
\begin{equation}
[H,Q_\alpha^\pm]=[H,Y]=[H,Z_{\alpha\beta}]=[Y,Z_{\alpha\beta}]=0.
\label{comm1}\end{equation}
As we see, this Hamiltonian differs from the one emerging in the parasupersymmetric model
by the insertion of one more copy of the middle subhamiltonian.

Matrices
$$
J_0:=\tfrac{1}{c}Z_{11},\quad J_+:=\tfrac{1}{2c}Z_{12}\quad J_-:=\tfrac{1}{2c}Z_{21}
$$
form the $sl(2)$ algebra, $[J_0, J_\pm]=\pm 2J_\pm,\; [J_+,J_-]=J_0.$
They have the following commutation relations with other operators
\begin{equation}
[Q_\alpha^\pm, J_0]=\pm Q_\alpha^\pm,\quad [Q_1^\pm, J_\pm]=\pm Q_2^\pm,
\quad [ Q_2^\mp, J_\pm]=\mp Q_1^\mp,
\label{comm2}\end{equation}
with vanishing other commutators $[Q_1^\mp, J_\pm]=[Q_2^\pm, J_\pm]=0.$
The described WSQM model was originally written with the help of creation and annihilation
operators for two fermions, which can be realized by $4\times4$ matrices. Here we omit
this representation and refer for its details to  \cite{Sweak}.

In Fig. 1 we presented all possible types of spectra for the WSQM model Hamiltonian.
Simultaneously, these describe the spectra for the PSQM Hamiltonian, as they were
listed in \cite{RS}. The latter emerge after deletion of
one of the middle towers of states (say, for $h_3$). If one deletes both middle
towers of states, leaving only the left and right extreme ones, then one comes to the $n=2$
HSQM Hamiltonian spectra.

\begin{figure}
\centering
\includegraphics[width=120mm]{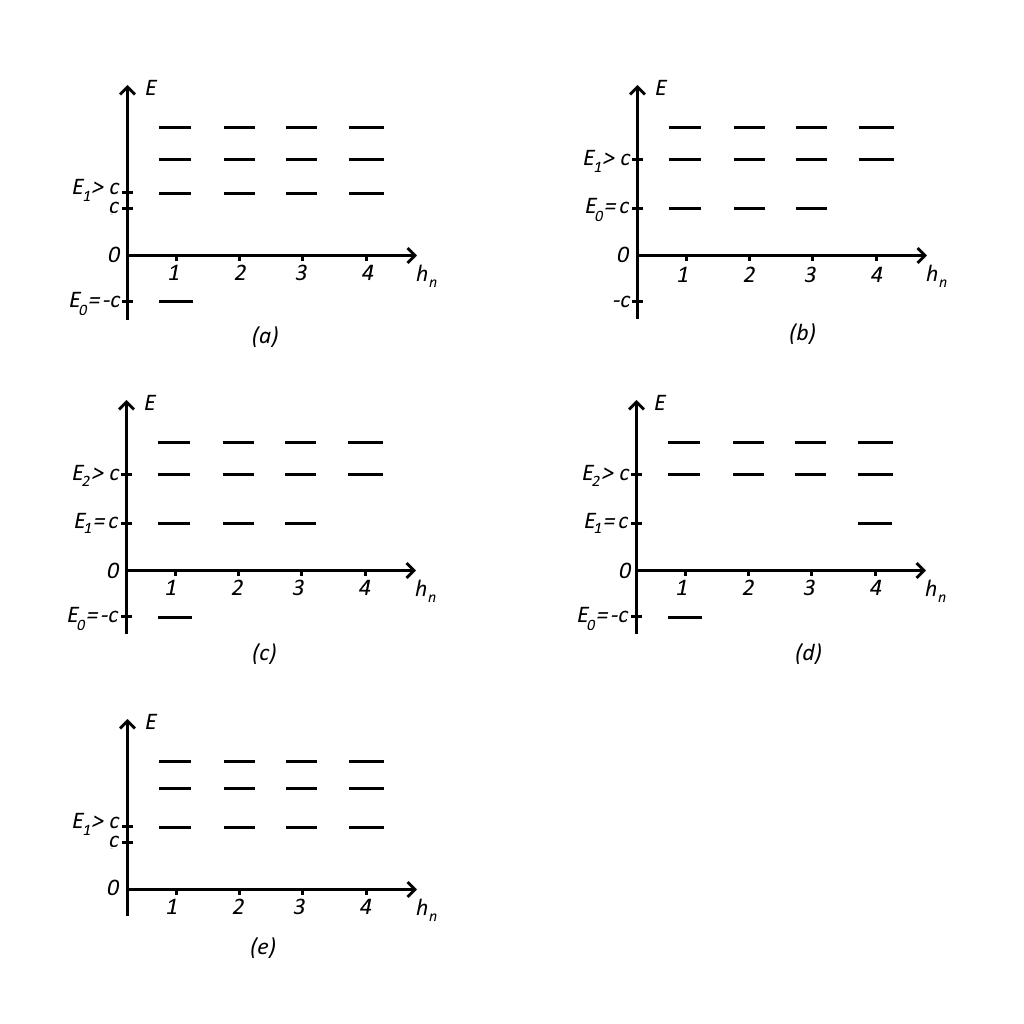}
\vspace{-11mm}
\caption{Forms of the spectra for the WSQM Hamiltonian.}
\label{figKP}
\end{figure}

One of the goals of the paper \cite{AIS} was to analyze the structure of Witten index for HSQM models.
It was shown that for $n>1$ this index has drastically different properties  from the standard case $n=1$.
In particular, it starts to depend on the fugacity used for computation of this index.
The notion of superconformal index was introduced in \cite{Rom1,KMMR}. It is an analogue of the Witten
index for an unusual realization of the supersymmetry algebra similar to the ``weak'' supersymmetry
case described above. This superconformal index is defined for a distinguished pair of supercharges and
depends on the fugacities introduced for all symmetry generators commuting with each other and with
these supercharges. Its important property is that it counts not the vacua,
but the so-called BPS states killed by the supercharges.

Let us compute for each $n=2$ HSQM spectrum pattern given in Fig. 1 the Witten index
\begin{equation}
I_W=Tr \left((-1)^Fe^{-\beta H}\right), \quad F:=f^-f^+,
\label{Wind}\end{equation}
where the fermionic charge $F$ is chosen in such a way, that the upper element of the column of
HSQM Hamiltonian eigenfunctions is identified with the bosonic sector.

In the WSQM case, we compute the superconformal index associated with supercharges $Q_1^\pm$:
\begin{equation}
I_{SCI}=Tr \left((-1)^Fe^{-\gamma\{Q_1^-,Q_1^+\}}e^{-\beta H}\right), \quad F=\tfrac{1}{c}Y+1.
\label{SCI}\end{equation}
Here the operator $F$ has eigenvalues $(0, 1,1,2)$, so that the upper and lower items
of the column of Hamiltonian eigenfunctions belong to the bosonic sector and the two
middle ones---to the fermionic sector.
This index formally contains two chemical potentials $\gamma$ and $\beta$,
but for the same reason as in the standard SQM case, dependence on $\gamma$
is actually absent. This happens because for non-zero eigenvalues of the
operator $\{Q_1^-,Q_1^+\}$ the $(-1)^F$ sign alternating factor forces to cancel equal
contributions. Only zero modes of the supercharges $Q_1^\pm \psi(x)=0$ (``BPS states'')
may give a contribution to this index.

The structure of the lowest energy levels depends on the normalizability of zero modes
of the operators $A^\pm_\alpha$,
$$
A^\pm_\alpha \phi_\alpha^\pm(x)=0, \quad \phi_\alpha^\pm(x)=\exp\big(\pm \int^xv_\alpha(y)dy\big).
$$
These functions are written up to arbitrary multiplicative factors, which can be fixed in cases
when $\phi_\alpha^\pm(x)\in \textrm{L}^2(\R)$ have unit norm.
Let us denote $v_1(x)=f(x)+B(x),\, v_2(x)=f(x)-B(x)$.
Then the constraint \eqref{cons} can be easily resolved, which yields
$B(x)=(c-f'(x))/(2f(x))$
for arbitrary function $f(x)$. This simplifies the analysis of asymptotic
behaviour of $\phi_\alpha^\pm(x)$  for $x\to \pm \infty$.

1) Let $\phi_1^-(x)$ be a normalizable function. Then $\phi_1^+(x)$ is
automatically not normalizable. Suppose that $\phi_2^\pm(x)$ are not normalizable  as well.
This situation is depicted in Fig. 1a. The vacuum state is $\psi_{E_0}(x)=(\phi_1^-(x), 0,0,0)^t$
(or $\psi_{E_0}(x)=(\phi_1^-(x), 0)^t$ for HSQM) with the energy $E_0=-c$.
It satisfies the conditions $Q_1^\pm\psi_{E_0}(x)=0$ (or $Q^\pm\psi_{E_0}(x)=0$ for HSQM) and yields
$$
I_W=I_{SCI}=e^{\beta c}.
$$
Note that in this case the second energy level has $E_1>c$.

2) Now we assume that $\phi_1^-(x)$ is not normalizable. Then, for $c>0$, the function
$\phi_1^+(x)$ cannot be normalizable alone. In this case $h_2$ and $h_3$ should have the lowest
eigenvalue $E_0=-c$, whereas the lowest eigenvalue of isospectral to them operator $h_4$ should
be bigger than $c$, which is not possible.

3) Let now only $\phi_2^-(x)$ be normalizable.
This situation is depicted in Fig. 1b. The ground state is triply degenerate with $E_0=c$:
$\psi_{E_0}(x)\propto (0,0,\phi_2^-(x),0)^t,\, (0,\phi_2^-(x),0,0)^t,\, (A_1^+\phi_2^-(x),0,0,0)^t$.
However, only one fermionic state satisfies the needed equations $Q_1^\pm\psi_{BPS}(x)=0$,
$\psi_{BPS}(x)=(0,\phi_2^-(x), - \phi_2^-(x),0)^t$, which gives the only non-zero contribution to $I_{SCI}$.
For HSQM we have $\psi_{E_0}(x)=(A_1^+\phi_2^-(x),0)^t$ with $E_0=c$ and $Q^\pm \psi_{E_0}(x)=0$.
As a result, we have
$$
I_W=-I_{SCI}=e^{-\beta c}.
$$
The second energy level $E_1>c$ remains unknown in general.

4) Assume now that
only $\phi_2^+(x)$ is normalizable. Then, $\psi_{E_0}(x)\propto (0,0,0,\phi_2^-(x))^t$
with $E_0=c$. As already mentioned, by the transformations $v_j(x)\to -v_j(x), \, c\to -c$,
this model reduces to the situation 1) after reshuffling the subhamiltonians $h_j\to h_{5-j}$.
Therefore we drop this case.

5) Now we consider the cases when two zero modes  are simultaneously normalizable.
First possibility is  that both $\phi_\alpha^-(x)$ represent physical states.
This situation is depicted in Fig. 1c. The ground state is unique $\psi_{E_0}(x)=(\phi_1^-(x), 0,0,0)^t$,
$E_0=-c$, $Q^\pm_1\psi_{E_0}(x)=0$. The next energy level $E_1=c$ is triply degenerate
$\psi_{E_1}(x)\propto (0,0,\phi_2^-(x),0)^t,\, (0,\phi_2^-(x),0,0)^t,\, (A_1^+\phi_2^-(x),0,0,0)^t$.
However, there is only one BPS state formed from them $\psi_{BPS}(x)=(0,\phi_2^-(x), - \phi_2^-(x),0)^t$,
$Q_1^\pm\psi_{BPS}(x)=0$. As a result, we have two nonzero contributions both to $I_W$ and $I_{SCI}$:
$$
I_W=e^{\beta c}+e^{-\beta c}, \quad I_{SCI}=e^{\beta c}-e^{-\beta c}.
$$
For HSQM we have $\psi_{E_0}(x)=(\phi_1^-(x),0)^t$ with $E_0=-c$
and $\psi_{E_1}(x)=(A_1^+\phi_2^-(x),0))^t$ with $E_1=c$, $Q^\pm \psi_{E_{0,1}}(x)=0$.
The third energy level $E_2>c$ remains unknown.

6) Let  $\phi_1^-(x)$ and $\phi_2^+(x)$ be simultaneously normalizable. This case is presented in Fig. 1d.
The lowest eigenvalue state $\psi_{E_0}(x)=(\phi_1^-(x),0,0,0)^t$ with $E_0=-c$ and
the next energy state $\psi_{E_1}(x)=(0,0,0,\phi_2^+(x))^t$ with $E_1=c$ are both BPS states,
$Q_1^\pm\psi_{E_0}(x)=Q_1^\pm\psi_{E_1}(x)=0$. For HSQM we have similar lowest energy eigenfunctions.
As a result, we have two non-zero contributions to both $I_W$ and $I_{SCI}$, but with different signs:
$$
I_W=e^{\beta c}-e^{-\beta c}, \quad I_{SCI}=e^{\beta c}+e^{-\beta c}.
$$
The value of $E_2>c$ is not fixed.

7) If both $\phi_\alpha^+(x)$ represent physical states, then by the transformation
$v_j(x)\to -v_j(x), \, c\to -c$, and reshuffling the subhamiltonians, we return to the case 5).
The situation when $\phi_1^+(x)$ and $\phi_2^-(x)$ both are normalizable is not possible

8) Finally, let none of $\phi_\alpha^\pm(x)$ is normalizable. This case is depicted in Fig. 1e.
There are no BPS states for WSQM and supersymmetric vacua for HSQM. The ground state
with unknown energy $E_0>c$ is four times degenerate and
$$
I_W=I_{SCI}=0.
$$

So, we see that both indices $I_W$ for HSQM and $I_{SCI}$ for WSQM
contain equivalent information about the structure of lowest energy states
(an equivalence of these models was anticipated in \cite{Sweak}).
The change of signs in front of certain terms occurs
purely because of the flip of the fermionic and bosonic state tags. In a
recent paper \cite{Sind} Smilga computed $I_{SCI}$ (referred to there as the generalized
Witten index) for a one-dimensional model corresponding to the situation of Fig. 1c
and for a more complicated two-dimensional system with infinitely many BPS states. Here we have
considered all possible one-dimensional cases and related them to the results of \cite{RS} and \cite{AIS}.

The paper \cite{RS}  was a turning point for the author in changing the subject of
 research from quantum field theory to mathematical physics problems. As shown in
 \cite{RS,AISV}, simple natural restrictions for PSQM models to describe a spin 1 particle
 in external field result in the potentials for which the whole spectra can be found exactly.
The author decided to understand the hidden mechanism for that and to search for the
 most general univariate exactly solvable model in quantum mechanics. In short, it has led to the
theory of special functions. Special functions can be interpreted as the functions
associated with self-similar solutions of the chains of spectral transformation
like \eqref{chain}, which are related to completely integrable systems. This gives a constructive tool for
discovering new examples of special functions. Especially, this was useful for building new exactly
solvable potentials in one-dimensional quantum mechanics \cite{S95}.

 Surprisingly, in more than 15 years after
 computing Witten indices in HSQM (i.e., the disguised superconformal indices) in the work
 \cite{AIS},   the author was forced to
investigate superconformal indices in four-dimensional $\mathcal{N}=1$ supersymmetric
field theories per se \cite{SV}. The reason for that came from the fact that these indices
coincided with the elliptic hypergeometric integrals representing a new class of
special functions of hypergeometric type discovered by the author \cite{spi:umn}.
And the latter functions were found precisely in a hunt for a universal most general
exactly solvable Schr\"odinger equation inspired by \cite{RS}.

Superconformal index for four-dimensional $\mathcal{N}=1$ supersymmetric field theories
with the gauge group $G$ and flavor group $F$ is a substantially more complicated object
than $I_{SCI}$ we considered above \cite{Rom1,KMMR} (for a survey, see \cite{RR2016}).
The flat space-time symmetry
group $SU(2,2|1)$ is generated by $J_i,\, \overline{J}_i$ (Lorentz
rotations), $P_\mu, Q_{\alpha},\overline{Q}_{\dot\alpha}$ (supertranslations),
$K_\mu, S_{\alpha},\overline{S}_{\dot\alpha}$ (special superconformal transformations),
$H$ (dilations) and $R$ ($U(1)_R$-rotations). The key relation,
which is preserved when the theory is put on the curved manifold $S^3\times S^1$,
is defined by a distinguished pair of nilpotent supercharges, say $Q\propto \overline{Q}_{1 }$
and  $Q^{\dag}\propto {\overline S}_{1}$, $Q^2=(Q^{\dag})^2=0$, and has the form
\begin{equation}
\{Q,Q^{\dag}\}= 2{\mathcal H},\qquad \mathcal{H}=H-2\overline{J}_3-3R/2.
\label{RSUSY}\end{equation}
Then the generators $M_l\in (J_3, \mathcal{R}, F_k, \mathcal{H})$, where $\mathcal{R}= H-R/2$
and $F_k$ are the flavor group maximal torus generators,
form the maximal set of operators commuting with supercharges and with each other
$[Q, M_l]=[Q^{\dag},M_l]=[M_l,M_{l'}]=0.$
Then the superconformal index is defined as the trace
\begin{eqnarray}\nonumber
&& I(p,q,y_k) = Tr \Big( (-1)^{N_f}
p^{\mathcal{R}/2+J_3}q^{\mathcal{R}/2-J_3}
 \prod_{k}y_k^{F_k}e^{-\gamma {\mathcal H}}\Big),
\label{Ind}\end{eqnarray}
where $N_f$ is the fermion charge, and
$p,q,y_k,\gamma$ are group parameters (fugacities or chemical potentials).
Here the dilation operator $H$ plays the role of Hamiltonian, not the operator ${\mathcal H}$
standing on the right-hand side of \eqref{RSUSY}.
Because of the presence of different signs for contributions from the bosonic and fermionc states,
the index $I(p,q,y_k) $ can get contributions only from the BPS states
$Q|\psi\rangle=Q^{\dag}|\psi\rangle=\mathcal{H}|\psi\rangle=0$ and therefore it does not depend on $\gamma$.

It was found that the superconformal index for a free chiral superfield with zero $R$-charge
is equal to the elliptic gamma function
$$
I_{chir}=\prod_{j,k=0}^\infty \frac{1-y^{-1}p^{j+1}q^{k+1}}
{1-yp^jq^k}\equiv\Gamma(y;p,q), \quad |p|,|q|<1,
$$
where $y$ is the fugacity for the corresponding $U(1)$ flavour group.
For gauge theories the index is given by a matrix integral over the Haar measure of the gauge group $G$.
R\"omelsberger conjectured \cite{Rom2} that superconformal indices of the theories related by the Seiberg
duality are equal. The simplest case of such non-abelian electromagnetic duality was described in \cite{S94}.
Corresponding electric theory has the gauge group $G=SU(2)$ and the flavour group $SU(6)$. It contains one
vector supermultiplet in the adjoint representation of $G$ and one chiral multiplet described by the
fundamental representations of $G$ and $F$. The $R$-charge of the latter field is 1/3.
The magnetic theory is described by the Wess-Zumino type model of one chiral
superfield described the antisymmetric tensor representation of $SU(6)$ with the $R$-charge 2/3.
This means that in the strong coupling regime one has the $s$-confinement (i.e., chiral symmetry is
not broken).

Explicit computation of the corresponding indices was performed by Dolan and Osborn \cite{DO}.
Electric superconformal index has the form
\begin{equation}
I_E= \frac{(p;p)_\infty(q;q)_\infty}{4\pi i}
\int_\T\frac{\prod_{j=1}^6
\Gamma(t_jz;p,q)\Gamma(t_jz^{{- 1}};p,q)}{\Gamma(z^{2};p,q)\Gamma(z^{-2};p,q)}\frac{dz}{z},
\label{IE}\end{equation}
where $(q;q)_\infty=\prod_{n=1}^\infty(1-q^n)$. The variables $t_j=(pq)^{1/6}y_j$ are related to
the flavour group fugacities $y_j$ and the $R$-charge of the chiral superfield.
They satisfy the balancing condition $\prod_{j=1}^6t_j=pq$, which is a direct consequence
of the constraint $\prod_{j=1}^6y_j=1$ for the maximal torus fugacities of the $SU(6)$ group.
The factor standing in front of the integral over unit circle $\T$ and the denominator of the integrand
describe the contribution from gluons and gluinos. The numerator of the integrand is
determined by the matter field contributions.

 Magnetic superconformal index is given by the following product of elliptic gamma functions
\begin{equation}
I_M=\prod_{1\leq j<k\leq6}\Gamma(t_jt_k;p,q).
\label{IM}\end{equation}
As shown in \cite{DO}, the R\"omelsberger conjecture holds true
in this case because of the Theorem  proven by the author in 2000 \cite{spi:umn}.
Namely, the exact integration formula for the elliptic beta integral \eqref{IE}
was established in  \cite{spi:umn}, which is identical with the equality $I_E  = I_M$ for $|p|, |q|, |t_j|<1$.
Thus the BPS state sectors in very differently  looking electric and magnetic
theories completely coincide.

The elliptic beta integral is the top univariate exactly computable
integral generalizing the Gaussian integral,
the Euler beta function and many other integrals serving as the orthogonality or biorthogonality
measures for some polynomials or rational functions. It comprises also the Newton's binomial
theorem and its various extensions. From the general theory of special functions point of view this is a
key element for the whole theory of transcendental elliptic hypergeometric functions unifying the standard
hypergeometric functions, their $q$-analogues and elliptic functions.

As to the mathematical physics applications, the elliptic hypergeometric functions emerged
in the theory of integrable many-body problems of the Ruijsenaars type and in the
solutions of the Yang-Baxter equation. The elliptic beta integral evaluation provides the most general
known solution of the star-triangle relation serving as a key to solvability  of two-dimensional
statistical mechanics systems of the Ising type. Simply speaking, the Seiberg duality is equivalent
to integrability of such systems. See \cite{rev} for a brief survey of these applications.

The process of evaluating elliptic beta integrals on root systems (multidimensional
analogues of the above formula) acquires an intriguing physical meaning. Specifically, it
describes a transition from the weak to strong coupling regime and the exact computability
serves as a criterion for $s$-confinement of the associated field theories.
Of course, it was impossible to foresee such impressive developments
from plays with the variations (``weakening'') of supersymmetric quantum mechanics
in \cite{RS,AISV,AIS,Sweak}. Still, they may be considered as some
rudimentary predecessors of these achievements.

\smallskip

{\bf Acknowledgements.}
First and foremost, I would like to sincerely thank Valery Rubakov for the enthusiastic response back in 1987
to my rather vague ideas about parastatistics, which led to the joint paper \cite{RS}.
Without such a response my scientific trajectory could have been quite different.
Also, I am indebted to Andrei Smilga for drawing attention to his WSQM model and useful discussions.
This study has been partially supported by the Russian Science Foundation (grant 24-21-00466).


\begin{thebibliography}{000000}

\bibitem{Wi81} E. Witten, {\em Dynamical breaking of supersymmetry}, Nucl. Phys.
{\bf B 188} (1981), no. 3, 513--554.

\bibitem{RS} V. A. Rubakov and V. P. Spiridonov, {\em Parasupersymmetric quantum
mechanics}, Mod. Phys. Lett. {\bf A 3} (1988), 1337--1347.

\bibitem{Inf} L. Infeld, {\em On a new treatment of some eigenvalue
problems}, Phys. Rev. {\bf 59} (1941), 737--747.


\bibitem{IH} L. Infeld and T. E. Hull, {\em The factorization method},
Rev. Mod. Phys. {\bf 23} (1951), 21--68.


\bibitem{AISV} A. A. Andrianov,  M. V. Ioffe, V. P. Spiridonov, and  L. Vinet,
{\em Parasupersymmetry and truncated supersymmetry in quantum mechanics},
Phys. Lett. {\bf B 272} (1991), 297--304.

\bibitem{AIS} A. A. Andrianov,  M. V. Ioffe, and V. P. Spiridonov,
{\em Higher-derivative supersymmetry and the Witten index},
Phys. Lett. {\bf A 174} (1993), 273--279.

\bibitem{AI}  A. A. Andrianov and M. V. Ioffe, {\em Nonlinear supersymmetric
quantum mechanics: concepts and realizations}, J. Phys. A: Meth. Theor.
{\bf 45} (2012), 503001.

\bibitem{Sweak} A. V. Smilga, {\em Weak supersymmetry}, Phys. Lett. {\bf B 585} (2004), 173--179.

\bibitem{Rom1} C. R\"omelsberger, {\em Counting chiral
primaries in ${\mathcal N}=1$, $d=4$ superconformal field theories},
Nucl. Phys. {\bf B 747} (2006), 329--353.


\bibitem{KMMR} J. Kinney, J. M. Maldacena, S. Minwalla, S. Raju,
{\em An index for 4 dimensional super conformal theories},
Commun. Math. Phys. {\bf 275} (2007), 209--254.

\bibitem{Sind} A. Smilga, {\em Witten index for weak supersymmetric systems: Invariance
under deformations}, Int. J. Modern Phys. {\bf A 37} (2022), no. 18, 2250118.

\bibitem{S95} V. P. Spiridonov, {\em The factorization method, self-similar
potentials and quantum algebras}, in: {\it Special functions-2000},
Kluwer, Dordrecht (2001), pp. 335--364; arXiv:hep-th/0302046.

\bibitem{SV}
V.~P.~Spiridonov and G.~S.~Vartanov, \textit{Elliptic hypergeometry
of supersymmetric dualities}, Commun. Math. Phys. {\bf 304} (2011), 797--874.

\bibitem{spi:umn}
V. P. Spiridonov, {\em On the elliptic beta function},
%Uspekhi Mat. Nauk {\bf 56} (1) (2001), 181--182
Russian Math. Surveys {\bf 56}:1 (2001), 185--186.


\bibitem{RR2016}
L. Rastelli and S. S. Razamat, {\em The supersymmetric index
in four dimensions}, J. Phys. A: Math. and Theor. {\bf 50} (2017), 443013.

\bibitem{Rom2} C. R\"omelsberger,
\textit{Calculating the superconformal index and Seiberg duality},
arXiv:0707.3702 [hep-th].

\bibitem{S94} N. Seiberg, \textit{Exact results on the space of vacua of four-dimensional SUSY gauge
theories}, Phys. Rev. {\bf D 49} (1994), 6857--6863

\bibitem{DO}
 F. A. Dolan and H. Osborn, {\em Applications of the superconformal index for protected
 operators and $q$-hypergeometric identities to ${\mathcal N}=1$ dual theories},
Nucl. Phys. {\bf B 818} (2009), 137--178.

\bibitem{rev}
V. P. Spiridonov, {\em Elliptic hypergeometric functions},
a complementary chapter to the Russian edition of the book
G. E. Andrews, R. Askey, and R. Roy, {\em Special functions}
(Cambridge Univ. Press, 1999),
Moscow, MCCME (2013), pp. 577--606; arXiv:0704.3099v2.

\end{thebibliography}
\end{document}